\RequirePackage{fix-cm}
\documentclass[smallextended]{svjour3}       
\smartqed  
\usepackage{graphicx}
\usepackage{mathptmx}      
\usepackage{amsmath}
\usepackage{multirow}
 \journalname{European Journal of Epidemiology}
\begin{document}

\title{Quantifying the under-reporting of genital warts cases
}


\author{David Mori\~na         \and
        Amanda Fern\'andez-Fontelo \and
        Alejandra Caba\~na \and
        Pedro Puig  \and
        Laura Monfil \and
        Maria Brotons \and
        Mireia Diaz
}


\institute{D. Mori\~na \at
              Barcelona Graduate School of Mathematics (BGSMath), Departament de Matem\`atiques, Universitat Aut\`onoma de Barcelona (UAB) \\
              Department of Econometrics, Statistics and Applied Economics, Riskcenter-IREA, Universitat de Barcelona (UB)
              \email{dmorina@mat.uab.cat}           
           \and
           A. Fern\'andez-Fontelo \at
              Chair of Statistics, School of Business and Economics, Humboldt-Universit\"at zu Berlin, Berlin, Germany
            \and
            A. Caba\~na \at
            Barcelona Graduate School of Mathematics (BGSMath), Departament de Matem\`atiques, Universitat Aut\`onoma de Barcelona (UAB) \and
            P. Puig \at
            Barcelona Graduate School of Mathematics (BGSMath), Departament de Matem\`atiques, Universitat Aut\`onoma de Barcelona (UAB) \and
            L. Monfil \at
            Unit of Infections and Cancer - Information and Interventions (UNIC - I\&I), Cancer Epidemiology Research Program (CERP), Catalan Institute of Oncology (ICO)-IDIBELL, L'Hospitalet de Llobregat (Barcelona), Spain \and
            M. Brotons \at
            Unit of Infections and Cancer - Information and Interventions (UNIC - I\&I), Cancer Epidemiology Research Program (CERP), Catalan Institute of Oncology (ICO)-IDIBELL, L'Hospitalet de Llobregat (Barcelona), Spain \and
            M. Diaz \at
            Unit of Infections and Cancer - Information and Interventions (UNIC - I\&I), Cancer Epidemiology Research Program (CERP), Catalan Institute of Oncology (ICO)-IDIBELL, L'Hospitalet de Llobregat (Barcelona), Spain
}

\date{Received: date / Accepted: date}

\maketitle

\begin{abstract}
Genital warts are a common and highly contagious sexually transmitted disease. They have a large economic burden and affect several aspects of quality of life. Incidence data underestimate the real occurrence of genital warts because this infection is often under-reported, mostly due to their specific characteristics such as the asymptomatic course. Genital warts cases for the analysis were obtained from the catalan public health system database (SIDIAP) for the period 2009-2016, covering 74\% of the Catalan population. People under 15 and over 94 years old were excluded from the analysis as the incidence of genital warts in this population is negligible. This work introduces a time series model based on a mixture of two distributions, capable of detecting the presence of under-reporting in the data. In order to identify potential differences in the magnitude of the under-reporting issue depending on sex and age, these covariates were included in the model. This work shows that only about 80\% in average of genital warts incidence in Catalunya in the period 2009-2016 was registered, although the frequency of under-reporting has been decreasing over the study period. It can also be seen that the under-reported issue has a deeper impact on women over 30 years old. The registered incidence in the Catalan public health system is underestimating the real burden in almost 10,000 cases in Catalunya, around 23\% of the registered cases. The total annual cost in Catalunya is underestimated in at least about 10 million Euros respect the 54 million Euros annually devoted to genital warts in Catalunya, representing 0.4\% of the total budget of the public health system.
\keywords{genital warts \and estimation \and HPV \and under-reporting \and time series}
\end{abstract}

\section{Introduction}\label{intro}
Health information systems are essential to ensure the safety and quality of health care and improve adherence to clinical practice guidelines, but they are also a very powerful tool concerning resources management and control, decision making, and effective and efficient planning of prevention and control interventions\cite{Groseclose2017,Ford2012}⁠. However, the incompleteness and inaccuracy of the information is common in this type of registries and can lead to problems at a clinical level, but also at a population level such as the underestimation of some diseases. In Catalunya (Spain), the Information System for Research in Primary Care (SIDIAP) was launched in 2010 with the integration of data from the clinical work station of primary care (ECAP) of the Catalan Health Institute (ICS), which started in 1998, and other complementary sources\cite{SIDIAP}⁠. The ICS is the main provider of health services in Catalunya and manages 283 out of 370 Primary Care Teams with a catchment of 5,564,292 people, approximately 74\% of the Catalan population\footnote{http://ics.gencat.cat/es/lics/}. Nevertheless, it is reasonable to assume that the incidence of genital warts (GW) will be very similar among the Catalan population not covered by ICS.
In the particular case of sexually transmitted diseases, it is even more important to have reliable information due to their remarkable morbidity, and therefore, the importance of controlling trends over time and priority setting (see \cite{McCormack2019}⁠⁠ for a comprehensive discussion focused on developing countries). GW are a common and highly contagious sexually transmitted disease in Catalunya (in 2016 the incidence was about 107 cases per 100,000 women and 139 cases per 100,000 men\cite{Brotons2018}⁠) caused by a subset of HPV types, with the most common being genotypes 6 and 11. They are usually benign, or non-cancerous, skin growths that develop on the genital area. However, they have an important negative impact on the health service and the individual, in addition to have a large economic burden and affect several aspects of quality of life\cite{Woodhall2011,Senecal2011,Castellsague2009}⁠. It is well known that incidence data underestimate, to some degree, the real occurrence of genital warts because this infection is often under-reported, mostly due to their specific characteristics such as the asymptomatic course of the disease\cite{Hsueh2009}⁠. Further, the SIDIAP database only includes data from the public healthcare sector and around 28\% of the general population in Catalunya have a double health insurance coverage, public and private\cite{ESCA2017}, so this fact can also explain why GW incidence rates are underestimated⁠, although this source of under-reporting cannot be detected by the proposed model as we only have data from the public health system.
There has been a growing interest in the past recent years to deal with data that are only partially registered or under-reported in the biomedical literature\cite{Fernandez-Fontelo2016,Bernard2014,Alfonso2015,Rosenman2006,Arendt2013,FernandezFontelo2019}⁠. Most of these previous works deal with discrete-valued time series, whereas this paper is focused on the incidence of a disease, which should be treated as a continuous-valued time series. Therefore, the aim of this work is to quantify the under-reporting of genital warts cases in Catalunya and the reconstruction of the actual incidence in the period 2009-2016 on the basis of the mixture model described in Section~\ref{methods}.

\section{Methods}\label{methods}
\subsection{Population and incidence estimation}
The study population included all residents in Catalunya assigned to an ICS primary care center (74\% of the Catalan population). Monthly GW incident cases for the analysis were obtained from the SIDIAP database for the period 2009-2016. Episodes of GW were classified as incident if they were preceded by at least 12-month period without any episode. People under 15 and over 94 years old were excluded from the analysis as the incidence of GW in this population is negligible (averages of 0.24 cases and 0.22 x 100,000 individuals over the period of study respectively).

\subsection{Model}
Consider $X_t$ the series of real GW incidence, where $t=1, 2, \ldots$ is the time, following a normal distribution with mean $\mu$ and variance $\sigma^2$. In our setting, this process cannot be directly observed, and all we can see is a part of it, expressed as 
\begin{equation}\label{eq:model}
    Y_t=\left\{
                \begin{array}{ll}
                  X_t \text{ with probability } 1-\omega_t \\
                  q \cdot X_t \text{ with probability } \omega_t
                \end{array}
              \right.
\end{equation}

The series $Y_t$ represents the registered values corresponding to GW incidence in the part of Catalunya covered by ICS. According to Eq.~\ref{eq:model}, the registered observations series $Y_t$ is a mixture of two normally distributed random variables $Y_t=(1-\omega_t)\cdot Y_{1t} + w_t \cdot Y_{2t}$, where $Y_{1t}$ coincides with the unobserved process and $Y_{2t}$ is a normal random variable with mean $\mu$ and variance $\sigma^2$. The parameter $\omega_t$ is modeled as $\log(\omega_t) = \alpha_0 + \alpha_1 \cdot t$ and can be interpreted as the frequency of under-reporting at a time $t$, while $q$ can be interpreted as the intensity of such under-reporting, taking a value between 0 and 1. When $q=0$ the observed incidence is $Y_t=0$ and when $q=1$ there is no under-reporting. A value of $\omega_t$ equal to 0 indicates that the observed value at time $t$ is not under-reported, and a value of $\omega_t$ equal to 1 means that under-reporting is for sure happening. In order to detect potential differences in GW incidence depending on sex (men and women) and age (16-29 and 30-94), these covariates were included in the model, so the mean of the observed process $Y_{1t}$ was modeled as $\mu_{1t}=\beta_0 + \beta_1 \cdot t + \beta_2 \cdot a + \beta_3 \cdot s + \beta_4 \cdot a*s$ (where $a$ is the age, $s$ is the sex and $a*s$ is the interaction between age and sex). Similarly, the average of the second component  can be recovered as $\mu_{2t}=q \cdot \left(\beta_0 + \beta_1 \cdot t + \beta_2 \cdot a + \beta_3 \cdot s + \beta_4 \cdot a*s \right)$.
After fitting the previous model and performing residuals examination, a seasonal behavior with period 3 months was observed. Hence the model was updated by including the following trigonometric function to reflect this periodic behavior:
\begin{equation*}
 f(t) = \beta_5 \cdot \sin \left(\frac{2 \cdot \pi \cdot t}{3} \right) + \beta_6 \cdot \cos \left(\frac{2 \cdot \pi \cdot t}{3} \right),
\end{equation*}
on the terms $\mu_{1t}$ and $\mu_{2t}$.

Other similar models were considered and the best fitting one, according to the validation process described in Section~\ref{validation}, was chosen. In particular, as coefficients $\beta_1$ and $\beta_6$ are not significant, models without linear trend and with only one periodicity term were considered but the resulting validations were not satisfactory. 
The estimates and their associated standard errors were obtained by maximizing the likelihood function using the \textit{nlm} procedure in R\cite{RCoreTeam2019}⁠. The \textit{R} package \textit{mixtools}\cite{Benaglia2009a}⁠ was used to obtain proper initial values for the maximization algorithm. All the data and code used are available as supplementary material. If the main focus is not on quantifying the under-reporting issue, an alternative approach to analyze these data might be a hierarchical generalized linear model with random effects\cite{Lee1996}⁠, implemented in the \textit{R} package HGLMM\cite{Molas2011}⁠.
By means of this methodology the most likely unobserved real GW incidence process is reconstructed using the components of the estimated mixture, provided by the output of the \textit{mixtools} procedure.

\subsection{Validation}\label{validation}
\begin{figure}[ht!]
  \includegraphics[width=0.95\textwidth]{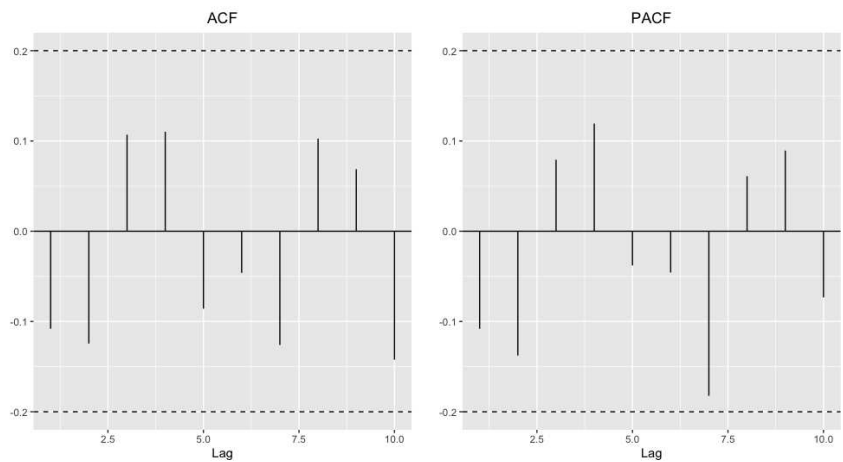}
\caption{Auto-correlations and partial auto-correlations of the model residuals.}
\label{fig:1}       
\end{figure}

The model has been validated by analyzing its residuals. Figure~\ref{fig:1} shows that they behave like white noise as expected and that there are no significant auto-correlations that should be accounted for. The residuals $r_t$ have been estimated as
\begin{equation*}
 \hat{r_t} = Y_t - \left(\hat{\omega_t} \cdot \hat{q} \cdot \left(\hat{\beta_0}+\hat{\beta_1} \cdot t + \hat{\beta_2} \cdot a + \hat{\beta_3} \cdot s + \hat{\beta_4} \cdot a*s \right) + (1-\hat{\omega_t}) \cdot \left( \hat{\beta_0}+\hat{\beta_1} \cdot t + \hat{\beta_2} \cdot a + \hat{\beta_3} \cdot s + \hat{\beta_4} \cdot a*s \right) \right)
\end{equation*}

where $Y_t$ is the total observed GW incidence at time $t$, and the letters with hat indicate the estimated parameters.

\section{Results}\label{results}
Our analysis estimates that, globally, only around 80\% of actual GW incidence was registered in the SIDIAP database in the period 2009-2016. For women over 30 years old, the monthly average registered incidence is 3.9 cases per 100,000 women, while the estimated monthly incidence is 4.9 cases per 100,000 women, 24.9\% higher. On males over 30 years old, the registered series has a monthly average of 5.9 cases per 100,000 men for 7.1 cases per 100,000 men on the reconstructed series, 21.8\% higher. Regarding males under 30 years old, the reconstructed series is 13.3\% higher (monthly averages of 18.4 and 20.8 cases per 100,000 men for the registered and reconstructed processes respectively). For women under 30 years old, the monthly average registered incidence of GW in Catalunya is 19.0 per 100,000 women, while the reconstructed hidden process has an average of 23.0 cases per 100,000 women, about 21.0\% larger. This information is summarized in Table~\ref{tab:1} and described in more detail in the supplementary material (Table S1).
\begin{table}
\caption{Registered and estimated GW monthly average incidence (number of cases x 100,000 individuals) in the period 2009-2016.}
\label{tab:1}       
\begin{tabular}{lllll}
\hline\noalign{\smallskip}
Sex & Age & Incidence (registered) & Incidence (estimated) & Difference (\%)  \\
\noalign{\smallskip}\hline\noalign{\smallskip}
\multirow{3}{*}{Females} & 15-29   & 19.0 & 23.0 & 21.0\% \\
                         & 30-94   & 3.9  & 4.9  & 24.9\% \\
                         & Average & 6.8  & 8.4  & 23.2\% \\
\hline
\multirow{3}{*}{Males}   & 15-29   & 18.4 & 20.8 & 13.3\% \\
                         & 30-94   & 5.9  & 7.1  & 21.8\% \\
                         & Average & 8.3  & 9.8  & 18.3\% \\
\hline
Global                   &         & 7.6  & 9.1  & 19.9\%\\
\noalign{\smallskip}\hline
\end{tabular}
\end{table}

Table~\ref{tab:2} shows the estimated effect of the age and sex over the under-reporting issue. In particular, it can be seen that the GW incidence is higher among younger populations and men. It can also be noticed that a significant interaction between sex and age group is found, which can be interpreted as a distinguishable impact of sex on GW incidence depending on the age group.
\begin{table}
\caption{Parameter estimates.}
\label{tab:2}       
\begin{tabular}{lll}
\hline\noalign{\smallskip}
Covariate & Parameter & Estimate (95\% CI) \\
\noalign{\smallskip}\hline\noalign{\smallskip}
           & $\alpha_0$ & 2.99 (1.77, 4.20) \\
$t$        & $\alpha_1$ & -4.31 (-6.53, -2.09) \\
           & $\beta_0$  & 13.76 (7.11, 20.40) \\
$t$        & $\beta_1$  & 0.36 (-12.75, 13.46) \\
$Age$      & $\beta_2$  & -13.53 (-14.13, -12.92) \\
$Sex$      & $\beta_3$  & -1.60 (-2.24, -0.95) \\
$Age*Sex$  & $\beta_4$  & 3.25 (2.44, 4.06) \\
           & $\beta_5$  & 4.16 (0.44, 7.88) \\
           & $\beta_6$  & 0.52 (-5.59, 6.64) \\
           & $q$        & 0.75 (0.72, 0.77) \\
\noalign{\smallskip}\hline
\end{tabular}
\end{table}

\begin{figure}[ht!]
  \includegraphics[width=0.95\textwidth]{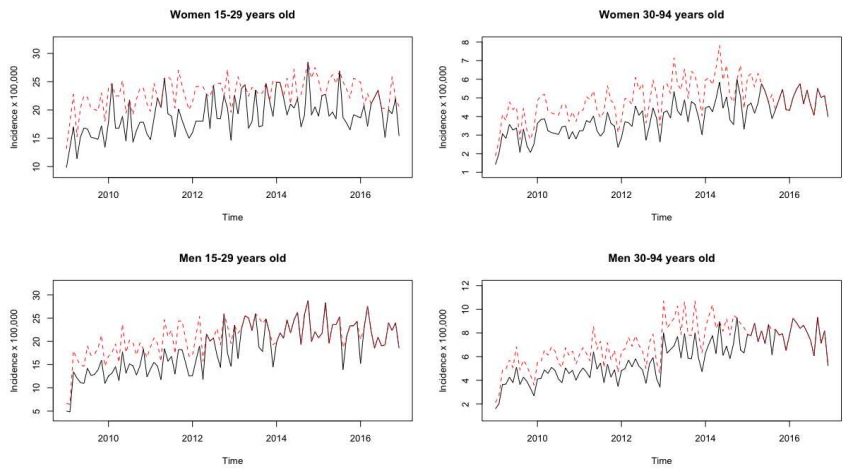}
\caption{Registered (solid black line) and estimated underlying series (dashed red line) for each of the considered sub-populations.}
\label{fig:2}       
\end{figure}

Figure~\ref{fig:2} shows the registered (solid black line) and reconstructed unobserved (dashed red line) processes for each of the considered sub-populations. Although this figure shows increasing trends for all series, they are not well explained by coefficient $\beta_1$, which is not significantly different from zero. Increasing trends are mainly explained by the significant coefficient $\alpha_1$, which leads to a decreasing frequency of under-reporting $\omega_t$.

The under-reporting frequency is about 95\% in 2009 ($\omega_1$) and around 21\% in 2016 ($\omega_t$). This is measured by parameter $\alpha_1$ in Eq~\ref{eq:model}, and should not be confused to overall under-reporting of the data, as its intensity (measured by parameter $q$ in the model) also plays a crucial role. For instance, all observations in a certain period of time could be slightly under-reported ($\omega=1$, $q$ near to 1), resulting in small differences between registered and estimated values or just a few observations might be under-reported ($\omega$ near to zero) but with a high intensity ($q$ near to zero), potentially resulting on large differences between registered and estimated values. Table~\ref{tab:3} shows the total number of GW cases registered in the SIDIAP in the period of study, the reconstructed values according to these registered cases and the projection over the whole Catalan population, assuming that the incidence on the area outside ICS coverage is the same.

\begin{table}
\caption{Registered, estimated and projected number of GW cases in Catalunya.}
\label{tab:3}       
\begin{tabular}{llllll}
\hline\noalign{\smallskip}
Sex & Age & \begin{tabular}[t]{@{}c@{}}SIDIAP\\(registered)\end{tabular} & \begin{tabular}[t]{@{}c@{}}SIDIAP\\(estimated)\end{tabular} & \begin{tabular}[t]{@{}c@{}}Catalunya\\(registered projection)\end{tabular} &  \begin{tabular}[t]{@{}c@{}}Catalunya\\(estimated projection)\end{tabular}  \\
\noalign{\smallskip}\hline\noalign{\smallskip}
\multirow{3}{*}{Females} & 15-29   & 8051  & 9769  & 10280 & 12460 \\
                         & 30-94   & 7625  & 9520  & 9062  & 11337 \\
                         & Total   & 15676 & 19289 & 19342 & 23797 \\
\hline
\multirow{3}{*}{Males}   & 15-29   & 7967  & 9097  & 10166 & 11584 \\
                         & 30-94   & 10774 & 13842 & 12914 & 16598 \\
                         & Total   & 18741 & 22939 & 23080 & 28182 \\
\hline
Global                   &         & 34417 & 42228 & 42422 & 51979 \\
\noalign{\smallskip}\hline
\end{tabular}
\end{table}

\section{Discussion}\label{discussion}
The results of this work show that in relative terms, the under-reporting issue has a deeper impact on people over 30 years old (where GW incidence is lower), especially among women. Nonetheless, the relative difference between registered and estimated annual averages range between 13.3\% and 24.9\%. It is also remarkable that the quality of SIDIAP register regarding GW in Catalunya has been significantly improving during the study period, as the frequency of under-reported observations has been decreasing over time.
Facing under-reported information from public health registers is very common in many situations, especially regarding potentially asymptomatic diseases like GW. The proposed methodology considers the potential under-reporting in continuous time series data in a very flexible way, estimating its frequency and intensity, and it is general enough to be appropriate in a wide range of real situations in the public health context. Additionally, the most likely non-observed process can be reconstructed on the basis of estimated posterior probabilities. Moreover, the GW data show that these models can deal with time-dependent under-reporting parameters, seasonal behavior, trends and also incorporate the effect of other factors by including covariates.
The described methodology opens a wide field for future research lines. In particular, if temporal correlations are found in the data, an appropriate model should take this structure into account.
One of the potential limitations of this study is that the database used included data from the public healthcare setting and not from the private sector. In Catalunya, it is estimated that 33\% of women and 25\% of men aged 15 to 44 years have a double health insurance coverage (i.e. the public health insurance and a private insurance plan)\cite{ESCA2017}⁠, so the rates estimated in our study are likely still underestimating the real incidence of GW. One of its strengths is that the same methodology (possibly with minor model modifications) could be used to analyze the frequency and intensity of potential under-reporting issues for any condition or setting in the absence of temporal dependence among the observations.
The GW incidence registered in SIDIAP is underestimating the real burden in almost 10,000 cases in Catalunya, around 23\% of the registered cases. The annual per person cost of GW was around 1000 Euros\cite{Castellsague2009}⁠, so the potential total annual cost is underestimated in at least about 10 million Euros respect the 54 million Euros devoted to GW in Catalunya annually, representing 0.4\% of the total budget of the Catalan Government intended for health\footnote{https://catsalut.gencat.cat/ca/coneix-catsalut/informacio-economica/pressupost/}, although about 2.8 million Euros would correspond to private insurances.
It is, therefore, clear that knowing the true burden of GW at the general population level is important for health policy makers, especially after the introduction of prophylactic vaccines against HPV in many countries, as it plays a crucial role in developing and evaluating prevention strategies\cite{Kjaer2008,Kostaras2019}⁠.

\begin{acknowledgements}
David Moriña acknowledges financial support from the Spanish Ministry of Economy and Competitiveness, through the Mar\'ia de Maeztu Programme for Units of Excellence in R\&D (MDM-2014-0445) and Fundaci\'on Santander Universidades. We acknowledge the SIDIAP, with special thanks to Maria Aragon for her help in data collection. This work has partial funding promoted by the Department of Health of the Generalitat de Catalunya for the execution of the project Monitorizaci\'on y evaluaci\'on del impacto de la introducción de nuevas estrategias preventivas del cancer de cuello de \'utero en Catalunya (reference 0599S/7613/2010). This work was partially funded by the Instituto de Salud Carlos III-ISCIII (Spanish Government) through the projects PIE16/00049, PI16/01254, PI16/01056, PI19/01118, R\'io Hortega CM15/00061 (Co-funded by FEDER funds / European Regional Development Fund. ERDF, a way to build Europe), Ag\`encia de Gesti\'o d’Ajuts Universitaris i de Recerca (2017SGR1718) and by grant RTI2018-096072-B-I00 from the Spanish Ministry of Science, Innovation and Universities. We thank CERCA Programme / Generalitat de Catalunya for institutional support. Authors declare no competing personal or financial interests in relation to this study. Institutional support: The Cancer Epidemiology Research Programme (with which L.M., M.B. and M.D. are affiliated) has received sponsorship for grants from Merck and GlaxoSmithKline. The funding sources had no role in the data collection, analysis or interpretation of the results.
\end{acknowledgements}

%
\section*{Conflict of interest}
The authors declare that they have no conflict of interest.

\bibliographystyle{spphys}       
\bibliography{morina_fernandez_cabana_puig_monfil_brotons_diaz}   

%
%

\end{document}